\documentstyle[12pt]{article}
\newcommand{\doe}{This work was supported by the Director, Office of Energy
                  Research, Division of Nuclear Physics of the Office of High
                  Energy and Nuclear Physics of the U.S. Department of Energy
                  under Contract No. DE-AC03-76SF00098.}

\begin{document}

\begin{titlepage}

\begin{flushright}
     {\large LBL-34156}
  \end{flushright}

\vskip 2\baselineskip
\renewcommand{\thefootnote}{\fnsymbol{footnote}}
\setcounter{footnote}{0}
\begin{center}
\mbox{}\\[5ex]
{\LARGE Perturbative Gluon Shadowing }\\
{\LARGE in Heavy Nuclei{\footnote{\doe}} }\\[5ex]
{\large\hspace{20pt}K.J. Eskola,$\hskip-0.1truecm^{1,2}$ Jianwei Qiu$^3$  and
Xin-Nian Wang$^1$}\\[2ex]
        {\small\em $^1$Nuclear Science Division, Mailstop 70A-3307,
                        Lawrence Berkeley Laboratory,}\\
        {\small\em University of California, Berkeley, CA 94720, USA}\\
        {\small\em $^2$Laboratory of High Energy Physics, P.O. Box
        9, (Siltavuorenpenger 20C)\\
        SF-00014 University of Helsinki, Finland.}\\
        {\small\em $^3$Department of Physics and Astronomy,
        Iowa State University, \\Ames, IA 50011, USA}\\[2ex]
        {\normalsize July 23, 1993}\\
        \mbox{}\\[5ex]

\end{center}

\begin{abstract}
\normalsize
\baselineskip=20pt
We study how much gluon shadowing can be perturbatively
generated through the modified QCD evolution in heavy nuclei. The
evolution of small-$x$ gluons is investigated within the
semiclassical approximation.
The method of characteristics is used to evaluate the shadowed
distributions in low-$Q$ and small-$x$ region.
In solving the modified evolution equation,
we  model in simultaneously fusions from
independent constituents and from the same constituent,
both in a proton and in
a large loosely bound nucleus of $A\sim 200$. In addition to the actual
distributions at small $x$, we study the ratios of the distributions at an
initial scale $Q_0 = 2$ GeV, and show that a strong nuclear
shadowing can follow from the modified QCD evolution.
\end{abstract}

\end{titlepage}
\baselineskip 18pt
The semihard gluonic subprocesses are expected to play an
essential role in the
formation of high energy densities in heavy ion collisions at collider
energies \cite{KLLE87,WangG91,Geiger91,Ranft93}. However,
there are many theoretical uncertainties in
modeling these QCD processes. One of the major ones, nuclear gluon shadowing,
comes from the unknown initial gluon distributions at small $x$.
Unlike for the quark and antiquark distributions, there are no direct
experimental data for the gluons in nuclei. Getting theoretical control over
the nuclear gluon shadowing is therefore a very urgent and important issue.
The purpose of this Letter is to study how much gluonic
shadowing is generated perturbatively through  the modified QCD evolution
\cite{GribovLR81,MuellerQ86} in heavy nuclei.

``Shadowing'' in the context of the deep inelastic
$lA$-scattering refers to the
measured depletion of the nuclear structure function $F_2^A$
at small $x_{\rm Bj}$, as
compared to $F_2$ of unbound nucleons \cite{EMC90}. The same kind
of depletion at small $x$ is expected to happen also in the nuclear gluon
distributions. During the recent years there have been many efforts
to explain the measured nuclear shadowing of quarks and antiquarks
\cite{NikolaevZ75}-\cite{Kumano93}
but for  gluons the situation is still inconclusive. Once the nuclear parton
distributions are known at an initial scale $Q_0$,
the QCD-evolution to larger
$Q$ can be computed \cite{MuellerQ86,Qiu87,Eskola92}.
The problem is how to get
input, theoretically or experimentally, for the nuclear gluon distributions at
$Q_0$,
and to understand the reliability of QCD-evolution for the proper range of
$x$- and $Q$-values.

Shadowing-phenomenon is also predicted to happen in protons. In this case,
``shadowing'' refers to the  depletion of the actual parton distributions,
and is caused  by the  fusions of overcrowding gluons at very small $x$.
This mechanism proceeds through perturbative QCD-evolution  as formulated in
\cite{GribovLR81,MuellerQ86}. It has been  shown by Collins and
Kwieci{\'n}ski that the singular
gluon distributions actually saturate due to the fusions
\cite{CollinsK90}.

In this Letter our basic idea is quite straightforward. We first compute  the
gluon shadowing in a proton, by using techniques introduced in
\cite{CollinsK90}
for solving the small-$x$ evolution including gluon recombination. Then we
apply the same mechanism of recombining gluons to heavy nuclei
and study to what extent {\it nuclear} shadowing is generated perturbatively
through the QCD-evolution at $Q_0=2$ GeV.

At small values of $x$, leading order QCD evolution equation predicts
that the number of gluons becomes extremely large.  It has been known
\cite{GribovLR81,MuellerQ86} that for sufficiently small
values of $x$ and/or of $Q^2$, the
total
transverse area occupied by the
gluons will be larger than the transverse area of a hadron, so that
the interaction between gluons can no longer be neglected.  Such gluon
recombination results in a modification of the QCD evolution equations.
In the limit of small-$x$ the modified QCD evolution equation
can be cast in the form \cite{GribovLR81,MuellerQ86}
\begin{equation} \hskip-0.3truecm
\partial_y\partial_t G(y,t) = cG(y,t) - \lambda\exp(-t-{\rm e}^t)[G(y,t)]^2,
\label{APMQ}
\end{equation}
where $y = \ln(1/x)$, $t = \ln[\ln(Q^2/\Lambda^2_{\rm QCD})]$, $G(y,t) =
xg(x,Q^2)$ and $c = 12/(11-2N_f/3)$ with $N_f$ the number of quark
flavors.

Strength of the gluon recombination is controlled by the
factor $\lambda$, originating from two possible sources. The two fusing
gluon ladders, which couple 4 gluons to 2 gluons, can arise either from
independent constituents of proton/nucleus or from the same one,
as discussed in  \cite{MuellerQ86,CollinsK90,KwiecinskiMSR90}.
We will refer to the former case as ``independent'' and to the latter as
``non-independent'' fusion. Since  recombinations from  both
sources happen simultaneously, we divide the
parameter $\lambda$ into two parts:
\begin{equation}
\lambda = \lambda_{\rm I}+\lambda_{\rm II},\label{lambda}
\end{equation}
where $\lambda_{\rm I}$ corresponds to the independent recombination and
$\lambda_{\rm II}$ to the non-independent one.

Let us first study the two sources of recombination within the models for
two-gluon densities given in ref. \cite{MuellerQ86}. In a proton, the strength
of the independent fusion then takes  the form
\begin{equation}
\lambda_{\rm I} = \frac{2}{3}\frac{1}{\pi R_p^2}\cdot
\frac{\pi^3c^2}{2\Lambda_{\rm QCD}^2}, \label{lambdaIp}
\end{equation}
where $R_p\sim 1$ fm is the radius of a proton.

The magnitude of the  non-independent fusion of the gluon ladders can be
estimated as
\begin{equation}
\lambda_{\rm II} \approx \frac{16}{81}\frac{1}{\pi (2/Q_i)^2}\cdot
\frac{\pi^3c^2}{2\Lambda_{\rm QCD}^2}, \label{lambdaIIp}
\end{equation}
where we have made a simplification by fixing the initial $x$ of the valence
quark to $x_i\sim 1$. We also approximate the scale of the initial
valence quark by $Q_i\sim 2$ GeV.

Let us then consider a large loosely bound nucleus.
Naturally, both types of fusions are still there but only for the independent
one an $A^{1/3}$-scaling arises. In this case
\begin{equation}
\lambda_{\rm I}^A = \frac{9}{8}\frac{A}{\pi R_A^2}\cdot
\frac{\pi^3c^2}{2\Lambda_{\rm QCD}^2},\label{lambdaIA}
\end{equation}
where the nucleus is taken to be a sphere with a sharp surface at $R_A =
1.12A^{1/3}$ fm. The strength of  the non-independent fusion remains the same
as in the case of a free proton: $\lambda_{\rm II}^A = \lambda_{\rm II}$.

It is interesting to notice how the relative contributions of the two types of
recombination will change when going from a proton to a nucleus of $A\sim 200$:
$\lambda_{\rm II}/\lambda_{\rm I}\approx 7.6$ and
$\lambda_{\rm II}^A/\lambda_{\rm I}^A\approx 1.0$.
Thus the non-independent
fusion is clearly dominant in a free proton while in a large nucleus the
contributions from both types are of the same order.
As a result, parton recombination is strongly enhanced in a heavy nucleus.

In order to solve Eq.~\ref{APMQ} exactly by integration, one would need the
initial distribution either at fixed $y_0$ or $t_0$ {\it and} the
derivatives along a boundary line $(y,t_0)$ or $(y_0,t)$,
respectively.
However, since the expression for the non-linear term in
Eq.~\ref{APMQ} is not valid for the regions where $x$ is large,
or where
both $x$ and $Q$ are very small,
the natural boundary condition at $x=1$ (or $y_0=0$)
is not suitable here.
In addition,
since we do not have sufficient information on other boundary lines,
we cannot solve Eq.~\ref{APMQ} by direct integration.
Instead, with the semiclassical approximation \cite{GribovLR81}, we
are going to adopt the idea introduced in
\cite{CollinsK90} to use the method of characteristics,
so that we can avoid the region $III$ (see discussion later).

The semiclassical approximation corresponds to
neglecting the second order derivative term,
$\partial_y\partial_t\ln(G)$, which leaves us with the evolution equation as
\begin{equation}
\partial_yz(y,t)\partial_tz(y,t) = c-\lambda\exp[-t-{\rm e}^t+z(y,t)],
\label{semicl}
\end{equation}
where $z(y,t) = \ln[G(y,t)]$. The above equation can
then be cast and solved in
the form of a set of characteristic equations as shown
in detail in \cite{CollinsK90}.

The evolution of gluon distribution in a proton and in a nucleus is similar,
so let us first consider the general idea, as illustrated in Fig.~1.
We divide the $(y,t)$-plane into three regions. In region $I$ for $y\le y_0
(x\ge x_0)$, we expect the traditional, non-corrected
Altarelli-Parisi (AP) evolution \cite{AltarelliP77} to hold down to scales
$Q\sim 1$ GeV. In region $II$ with $y\ge y_0 (x\le x_0)$, the evolution of the
gluon distribution is then  approximately given by Eq.~\ref{semicl}.
The initial values of the
gluon distribution and  its $t$- and $y$-derivatives are
determined numerically at $(y_0,t_{\rm min}\le t\le t_0)$ from the
singular gluon distribution of the CTEQ-collaboration
\cite{CTEQ92}, which is obtained with a lower initial $Q$-value to provide the
necessary $t$-dependence of the boundary condition at $y_0$, and a larger
$x$-cut to ensure the validity of AP evolution in region $I$.
The evolution in the region $II$
is finally terminated at $t=t_0$, corresponding to $Q_0=2$ GeV.
Notice that the characteristics approach the $t_0$-line from below;
the lowest scale we have to
go down to is about $Q=1.25$ GeV, corresponding to $t_{\rm min}$.
At these scales QCD perturbation theory should be still valid
in region $I$.
In region $III$ with extremely large $y$ (small $x$) and/or
small
$t$
 we do not expect our analysis to be valid anymore, since  the higher order
 terms in the evolution equation will become important.

Before performing the actual evolutions, we have to consider how to
choose the boundary $y_0(=-\ln(x_0))$ for a proton and a nucleus, and how to
conserve momentum.

For a proton, we assume that the recombinations start to be effective at
$x\sim x_0\sim 0.01$, which is consistent with
\cite{CollinsK90,KwiecinskiMSR90}. We can use the results from global fitting,
like CTEQ, to constrain $x_0$. In fact, we will see that with $x_0=0.01$ the
shadowed gluons deviate considerably from the CTEQ gluons only after
$x<0.001$,
so the choice for $x_0$ seems to be reasonable, and we do not expect
the results to be very sensitive to small changes of $x_0$.

As explained above, the gluon
recombination is strongly enhanced in heavy nuclei and it starts at somewhat
larger values of $x$ than in protons. The corresponding boundary line $x_0^A$
for a nucleus is approximately determined by the relative magnitude of the
evolution terms in Eq.~\ref{APMQ}: $G_A(x_0^A)\sim G(x_0)\lambda_A/\lambda$,
so that the relative contribution from the gluon fusion in a nucleus is about
the same as in a nucleon.
This gives $x_0^A\sim$ 0.05--0.1. This range of $x_0^A$ is also supported by
other studies \cite{Eskola92}.

Let the original fraction of momentum in gluons
be $f_0=\int_0^1 dx \,xg_{\rm CTEQ}(x,Q_0^2)$. In the case of a proton,
shadowing
in the region $II$ changes the gluonic momentum typically by less than a per
cent, which we can clearly neglect as a small overall change.

Perturbative shadowing reduces the gluonic momentum more
in a nucleus than in a
proton. Assumed that the momentum fraction of gluons is conserved, there must
be a corresponding enhancement in the region $II$. In addition to this,
we also
take into account a possible  momentum transfer from quarks and antiquarks to
the gluons.  Here we consider nuclei with
$A\sim 200$, for which we expect an overall increase in the fraction of the
momentum, $\epsilon_A$, to be only about 4\%
\cite{FrankfurtS88,CloseQR89,Eskola92}. We combine these two sources of the
momentum flow, which  results in  solving $a_A$ iteratively from
\begin{equation}
\int_0^{x_0^A}dx\, xg(x,Q_0^2)\bigg|_{C}
+a_A\int_{x_0^A}^1dx\, xg_{\rm CTEQ}(x,Q_0^2)
=f_0(1+\epsilon_A),
\label{momP}
\end{equation}
with the condition $C:\, g_A(x_0^A,Q_0^2) =  a_A g_{\rm CTEQ}(x_0^A,Q_0^2)$ on
the boundary. Typically, $a_A\sim10$\% for $A\sim 200$.

Let us now turn to the results, presented in Figs.~2.
In Fig.~2a, nucleon and effective nuclear gluon distributions
for a nucleus of $A=200$ are compared with the input CTEQ gluon distribution
at $Q_0=2$ GeV.
Notice the $\sim$20 \% uncertainty in the nuclear case resulting from
varying $x_0^A$
from 0.05 to 0.1. To demonstrate the formation of strong perturbative nuclear
shadowing, corresponding to the {\it relative} depletion of gluon distributions
in a nucleus, we plot the ratio $G_A(x,Q_0^2)/G(x,Q_0^2)$ in Fig.~2b.
Notice also that as $x$ decreases, gluon distribution in a proton
increases much faster, or shows the sign of saturation at a much
smaller $x$ than that in a nucleus.  Therefore, as shown
in Fig.~2b the ratio saturates only when
the gluons in a {\it proton} do so. Thus, saturation of the perturbative
nuclear shadowing reflects actually the behavior of the gluons in a proton. As
a main result, we conclude that, due to the enhanced gluon recombination in a
heavy nucleus, a $\sim$50\% nuclear shadowing
in small-$x$ region
is generated perturbatively
through the modified QCD evolution, accompanied by a $\sim$10 \% antishadowing
from the momentum conservation.

As seen more clearly in Fig.~2b, with $x_0^A=0.1$ there is a slight
deviation in the initial derivatives of the gluon distributions
determined from
the region $I$ as compared to what can be determined from
Eq.~\ref{semicl} from
the region $II$. This in turn is a reflection of an apparent fact that
eventually one cannot apply the small-$x$ approximation at too large $x$.
Letting $x_0^A\sim 0.1$, we are really pushing the small-$x$
evolution equation
to its limit; surely beyond this point the Eqs.~\ref{APMQ} and
\ref{semicl} cannot be applied without additional correction terms. However,
taking the initial conditions from the ``known'' region $I$, as we do,
should improve the analysis and reduce the
uncertainty in the small-$x$ region. From Fig.~2b it is seen that we
cannot make conclusive claims about the ``beginning'' of nuclear shadowing.
However, since the result with $x_0^A=0.1$ does not differ considerably from
the result with $x_0^A=0.05$, we believe our result shows the correct order of
magnitude of the perturbative shadowing at very small $x$.

It is clear that the absolute strengths of the $\lambda$'s
depend on the models
assumed for the two-gluon densities in a proton and in a nucleus. However, we
do not expect our qualitative results for the nuclear shadowing to change very
much with different details.  One may also question
what happens to nuclear shadowing, if the initial gluons diverge more
strongly(weakly) when $x\rightarrow 0$ than
CTEQ gluon distribution used here.
In that case, the
recombinations would be enhanced(suppressed) both in a proton and in a heavy
nucleus.
However,
the result for nuclear shadowing,
$G_A(x,Q_0^2)/G(x,Q_0^2)$,
is not extremely sensitive
to the small-$x$ behavior of the input gluon distribution because we use
only the part with $x>0.01$, which has been relatively well-tested
experimentally.
This question will be studied
in more detail elsewhere \cite{EskolaWQfut}.

We would like to comment briefly on the general consequences of our result for
perturbative nuclear shadowing. The semihard processes with typical scales
$Q\sim$ a few GeV involve $x\sim Q/\sqrt s$. In heavy ion collisions at $\sqrt
s=200$ GeV, the $x$'s
in the semihard processes will be larger than 0.01, so these processes
rather probe the onset of perturbative nuclear shadowing than the region of
saturation. On the contrary, in  collisions with $\sqrt s$ in the TeV range,
the semihard processes will happen at $x$'s typically smaller than $10^{-3}$,
and are therefore affected considerably more by the perturbative shadowing.
Examples of the possible effects on  minijet production can be found in
\cite{WangG91,Eskola92,Eskola91}. Other processes clearly suppressed
by the nuclear gluon shadowing at very high energies are heavy quark and their
bound state production. Also the production of total transverse energy and
energy density will be  suppressed, as compared to the predictions with
non-shadowed gluons \cite{WangG91,Eskola91}.  Through this,  the
thermalization of the possibly formed quark-gluon plasma is also slowed down,
in
which case the thermal electromagnetic signals are suppressed.
In order to make
precise predictions for these processes, nuclear shadowing has
to be studied at
scales $Q>2$ GeV. The scale dependence of nuclear gluon shadowing is an
interesting question to which we will return in the  future
\cite{EskolaWQfut}.

To conclude,  we have considered the perturbative aspects of the nuclear
modifications to the gluon distributions.  As we have shown here, a strong
nuclear shadowing is generated through the modified QCD evolution, and it may
well be the dominant mechanism for the small-$x$ modifications.
We emphasize that the use of the method of characteristics is necessary
to avoid the region $III$ where even the modified evolution equation
is not expected to be valid.
We feel we now
have more quantitative control over the nuclear gluon distribution at small
$x$, based on perturbative QCD. We believe this study could serve as an
interesting starting point for more detailed calculations of nuclear gluon
shadowing and its consequences in ultra-relativistic heavy ion collisions.

\bigskip
\bigskip

\noindent {\bf Acknowledgements.}
   We acknowledge helpful discussions with J. Owens, R. Vogt and S. Gavin.
KJE is grateful  to Emil Aaltonen foundation, Magnus Ehrnrooth
foundation  and Suomen Kulttuurirahasto for partial financial support.
JWQ thanks the Nuclear Science Division at Lawrence Berkeley
Laboratory for hospitality while part of this work was
completed.
This work was supported by the U.S. Department of Energy under
contract No. DE-AC03-76SF00098, and in part by the Grant
Nos. DE-FG02-87ER40371 and DE-FG02-92ER40730, and by the
Texas National Research Laboratory
Commission.

\bigskip
\bigskip
\bigskip
\bigskip
\bigskip
\noindent {\Large \bf Figure Captions}
\bigskip
\bigskip

\noindent {\large\bf Fig. 1.}
The evolution plane. In the region $I$
the traditional AP-equations are expected to be valid.
Both $x$- and $Q$-dependence in this region form
the initial
conditions for the evolution in the small-$x$ (large $y$) region $II$.
Examples
of the characteristics of the Eq.~\ref{semicl} in the region $II$ are shown.
\label{fig1}

\bigskip
\noindent {\large\bf Fig. 2.}
{\bf a.} The gluon distributions $xg(x,Q_0^2)$ at $Q_0=2$ GeV vs.~$x$. The
result for proton is labeled by $x_0$, and the results for $A\sim 200$ by
$x_0^A$, respectively. The CTEQ gluon distribution \cite{CTEQ92} is labeled by
``CTEQ''. {\bf b.}  The ratio $xg_A(x,Q_0^2)/xg(x,Q_0^2)$
of the shadowed gluon
distributions  vs. $x$, demonstrating a strong perturbative nuclear shadowing
in heavy nuclei.
\label{fig2}

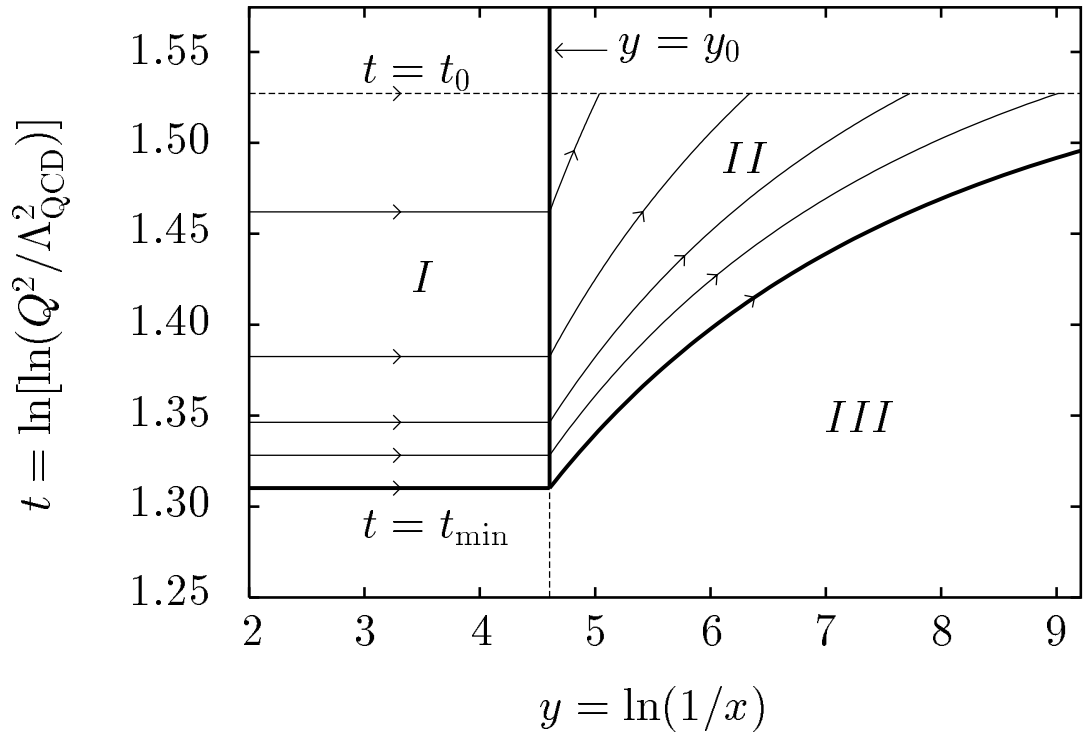
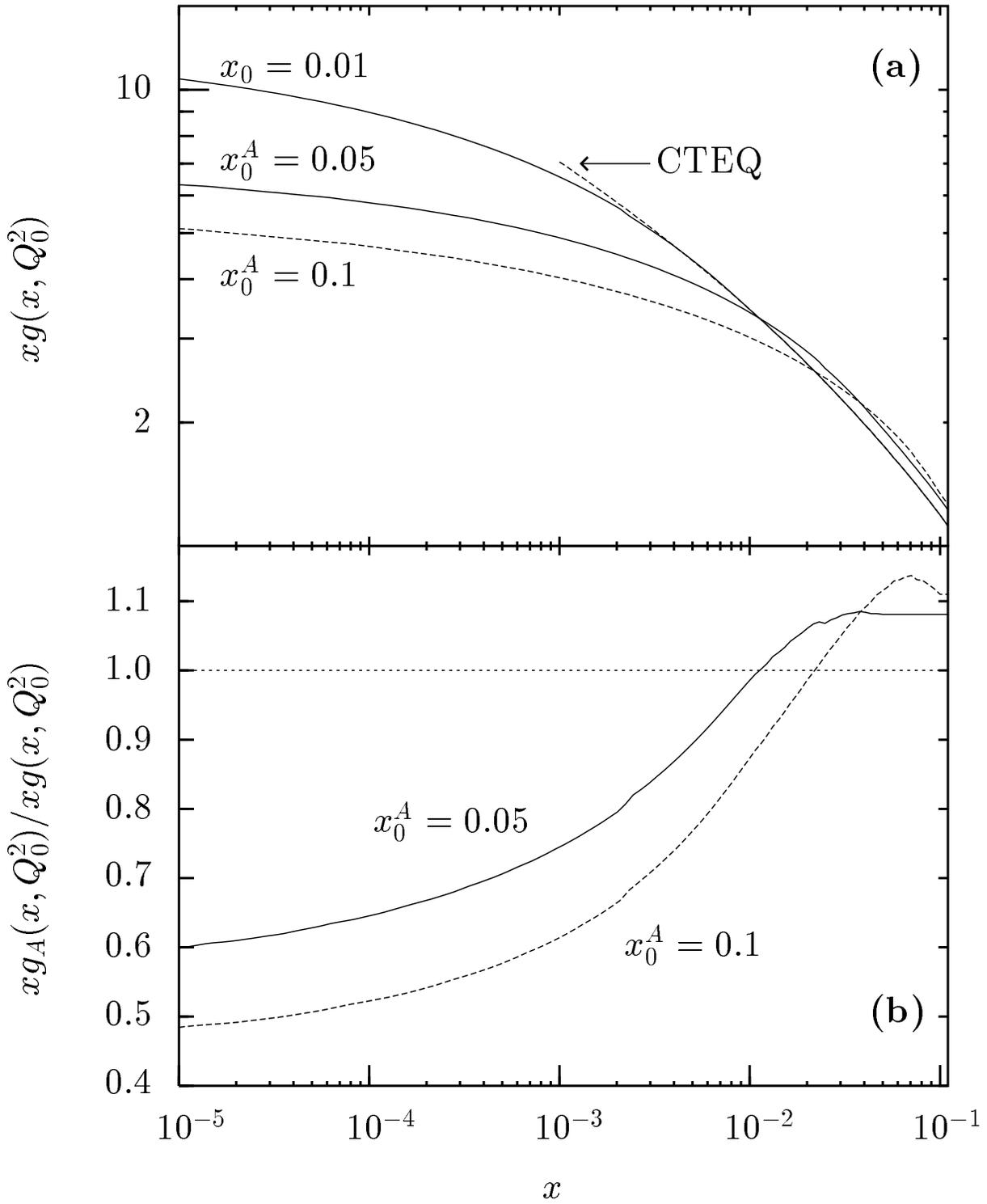
\end{document}